\documentclass[%
 aip,
 amsmath,amssymb,
reprint,%
]{revtex4-2}

\usepackage{graphicx}
\usepackage{dcolumn}
\usepackage{bm}
\usepackage{color}
\usepackage[mathlines]{lineno}

\usepackage[utf8]{inputenc}
\usepackage[T1]{fontenc}
\usepackage{mathptmx}

\DeclareMathAlphabet{\mathbfsf}{\encodingdefault}{\sfdefault}{bx}{sl}

\providecommand\nabla{\boldsymbol{\nabla}}
\providecommand\bcdot{\boldsymbol{\cdot}}

\newcommand{\hmin}{h_\textrm{min}}
\newcommand{\ls}{\ell_s}
\newcommand{\lsdim}{\lambda_s}
\newcommand{\tens}[1]{\mathbfsf{#1}}

\newcommand{\Ca}{\mbox{\textit{Ca}}} 	

\newcommand{\pd}[2] { \frac{\partial   #1}{\partial #2  } }

\begin{document}

\preprint{AIP/123-QED}

\title{The effect of wall slip on the dewetting of ultrathin films on solid substrates: Linear instability and second-order lubrication theory}

\author{A. Mart\'inez-Calvo}
    \email{amcalvo@ing.uc3m.es}
    \altaffiliation[]{Grupo de Mec\'anica de Fluidos,
    Departamento de Ingenier\'ia T\'ermica y de Fluidos,
    Universidad Carlos III de Madrid,
    Av.~Universidad 30,
    28911 Legan\'es (Madrid),
    Spain.}

\author{D. Moreno-Boza}%
    \email{damoreno@pa.uc3m.es}

\author{A. Sevilla}
    \email{asevilla@ing.uc3m.es}

\date{\today \, \, Published in Physics of Fluids} 

\begin{abstract}
The influence of wall slip on the instability of a non-wetting liquid film placed on a solid substrate is analyzed in the limit of negligible inertia. In particular, we focus on the stability properties of the film, comparing the performance of the three lubrication models available in the literature, namely the weak, intermediate and strong slip models, with the Stokes equations. Since none of the aforementioned leading-order lubrication models is shown to be able to predict the growth rate of perturbations for the whole range of slipping lengths, we develop a parabolic model able to accurately predict the linear dynamics of the film for arbitrary slip lengths.
\end{abstract}

\maketitle



\section{Introduction}\label{sec:intro}

Liquid films play a central role in many engineering applications, in biological and physiological processes, and in geophysics, to name a few. Apart from the fascination they hold for theoreticians for their rich dynamics, their great practical importance is evidenced by the existence of extensive reviews covering a large number of fundamental and applied studies~\citep{de1985wetting,Oron1997,Bonn2009,craster2009dynamics}, to which the reader is referred for a panoramic view of the subject. Many systems, such as plasmonic devices, or biofluids as in the case of tear film, involve coatings in the form of ultra-thin liquid films, which can be unstable if their height is below about $100~\mathrm{nm}$. Indeed, at these scales, the long-range van der Waals (vdW) forces can exceed the stabilizing surface tension force if the perturbation wavelength is above a certain cut-off~\citep{Scheludko1962,Vrij1966,Ruckenstein1974}. Many relevant applications involve the use of thin polymer films like silicone oils, which are known to experience a substantial slip when they flow over a solid impermeable substrate~\citep{deGennes1979,Brochard1992,Migler1993}. In these cases, the success of continuum mechanics to account for the observed phenomena requires the use of a \emph{slip length} $\lsdim$, defined as the distance to the wall at which the tangential velocity extrapolates to zero. 
The most commonly used wall-slip model is the linear boundary condition first derived by Navier~\citep{Navier1823}, and later on by Maxwell in the case of gases~\citep{Maxwell1879}. Slippage is important in many fields, ranging from microfluidics~\citep{Squires2005} to fracture and geophysical flows, or industrial flows involving polymer melt extruders, where slip-induced instabilities frequently occur~\citep{Denn2001} (see~\cite{Lauga2005,Neto2005} for a review).

\begin{figure*}
    \centering
    \includegraphics[width=420pt]{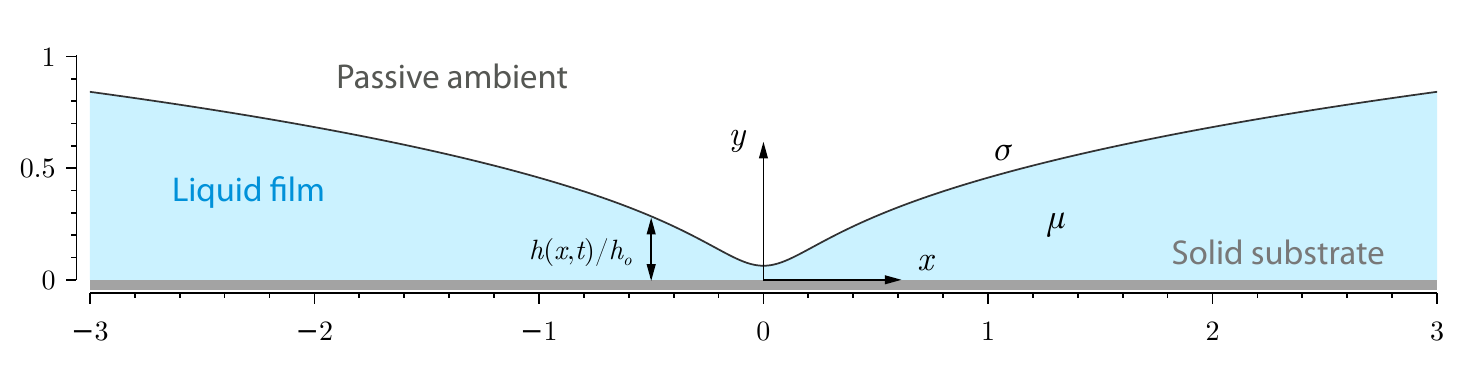}
    \caption{Schematic of the flow configuration with accompanying system of reference. Note that the complete streamwise domain, $-\pi/k < x < \pi/k $ is not fully represented in the sketch.}
    \label{fig:my_label}
\end{figure*}

We report the influence of wall slip on vdW-unstable liquid films, contemplating their linear stability. Since slippage is often found in the flow of polymeric films, the effects of inertia will be neglected herein, rendering the flow effectively Stokesian. More than forty years ago, de Gennes~\citep{deGennes1979} conjectured that the no-slip boundary condition at a rigid solid in contact with a semi-dilute or concentrated high-molecular-weight polymer solution may not apply. He argued that for any shear rate, under the \emph{ideal conditions} of a perfectly smooth non-adsorbing wall, the monomers do not create strong binds with the solid, and thus viscous forces in the liquid dominate over the friction with the substrate, leading to slip lengths as large as $\lsdim \sim 100~\mu\mathrm{m}$. Even larger slip lengths of $\lsdim \sim 1~\mathrm{cm}$ can be achieved by covering the substrate with a lubricant. Under \textit{semi-ideal conditions}, it was shown a decade later both theoretically~\citep{Brochard1992} and experimentally~\citep{Migler1993, Durliat1997} that the slip length in polymer-melt flows depends on the interaction between the monomers and the solid, and on the shear stress acting on the liquid. In particular, in this weakly grafted regime, usually known as the \textit{mushroom regime}, where bounded polymer chains do not overlap each other, it was shown by the latter authors that the slippage is strongly reduced ($\lsdim \sim 10~\mathrm{nm}$) due to the local binding of chains. However, as the shear rate increases, the slip velocity undergoes a sharp transition due to the \textit{coil-stretch transition} of the polymer blobs, whose elongation reduces the friction, inducing an increase in the slip length. Non-ideal conditions occur when the chains are bound to many locations along the substrate, as in the case of a strong \textit{brush}, and where the slippage is almost completely suppressed~\citep{Brochard1994}. Therefore, for thick films such that $\lsdim/h_o \ll 1$, where $h_o$ is the initial height of the film, a pressure gradient induces the standard semi-parabolic velocity profile. In contrast, for ultra-thin films where $\lsdim/h_o \gg 1$, a plug-flow velocity profile is expected. 

To date, most of the theoretical and numerical studies of the dynamics of non-wetting ultra-thin liquid films rely on leading-order lubrication models. However, it has been recently shown that the leading-order no-slip lubrication model fails to predict the near-rupture dynamics of ultra-thin films~\citep{morenobozaetal2020}. In the case of slipping films, there are three lubrication models available in the literature, namely the weak~\citep{Sharma1996}, intermediate~\cite{Munch2005} and strong~\cite{kargupta2004instability} slip models. In~\cite{kargupta2004instability}, the dispersion relation of a slipping vdW-unstable liquid film was deduced from the complete Navier-Stokes equations. However, it was not used to quantify the error of the two lubrication models available at that time. Here, we show that none of the leading-order models can cover the whole range of slip lengths, from the the no-slip limit, $\lsdim/h_o\ll 1$, to the opposite limit of a free film, $\lsdim/h_o \gg 1$. Indeed, the dispersion relations derived from the leading-order models fail at predicting the linear growth rates of small disturbances for arbitrary values of $\lsdim/h_o$. Specifically, we will show that the strong-slip model provides a reasonably accurate approximation for $\lsdim/h_o \gtrsim 10$ and any value of $h_o$, whereas the validity of the weak-slip model is restricted to $\lsdim/h_o \lesssim 0.1$ provided that $h_o/a \gtrsim 4$, where $a$ is the \emph{molecular length} defined below. Finally, the intermediate-slip model is only valid within a narrow range of $\lsdim/h_o$. It should be emphasized that these three models are still actively employed to describe the linear and nonlinear dynamics of the liquid film, for instance in a recent numerical study~\cite{Peschka2019}, where the intermediate lubrication model is used, to explain the instability in slipping dewetting rims~\citep{Fetzer2005,Baumchen2014}, or in numerical and experimental dip-coating studies with porous substrates, where the strong-slip model is used~\citep{Sathyanath2020}.

In view of these limitations, and given the present relevance of thin-film flows, the need to develop accurate lubrication approximations naturally arises. Thus, one of the main contributions of the present work is to develop a second-order lubrication model able to accurately predict the linear stability properties of the slipping film for arbitrary values of $\lsdim/h_o$. Our development is inspired by previous studies dealing with second-order lubrication theory in similar contexts like falling films~\citep{Ruyer1998,Ruyer2000,Ruyer2002,Kalliadasis2011} and axisymmetric liquid threads~\citep{GyC, Timmermans02,MartinezSevilla2018}.

The paper is organized as follows. In section~\ref{sec:stokes} we present an exact continuum formulation of the flow in terms of the Stokes equations subject to a Navier slip condition at the substrate. In section~\ref{sec:LO_lubrication} we present the three leading-order lubrication models developed in previous studies, followed by the development of a novel second-order lubrication approximation in section~\ref{sec:parabolic_main}. A detailed derivation of the parabolic lubrication model is presented in the Appendix.

\section{The Stokes equations}\label{sec:stokes}

The flow equations are made dimensionless using the unperturbed liquid film height $h_o$ as length scale, the vdW-induced velocity $A/(6 \pi \mu h_o^2)$ as velocity scale, their ratio $6 \pi \mu h_o^3/A$ as time scale, and the characteristic disjoining pressure $A/(6 \pi h_o^3)$ as pressure scale, where $\mu$ is the liquid viscosity and $A$ is the Hamaker constant~\citep{Hamaker1937}. Neglecting the inertia of the liquid, the flow is governed by the Stokes equations,
\begin{equation}\label{eq:continuity_momentum}
\nabla \bcdot \bm{u} = 0, \quad \text{and} \quad 
\bm{0} = - \nabla \phi + \nabla \bcdot \tens{T},
\end{equation}
where, $\bm{u} = (u,v)$, is the two-dimensional fluid velocity field in Cartesian coordinates $(x,y)$, $\tens{T} = - p\tens{I} + \nabla \bm{u} + (\nabla \bm{u})^{\rm{T}} $ is the stress tensor of an incompressible Newtonian fluid, $p$ is the pressure field, and $\phi =  h^{-3}$ is the dimensionless vdW potential. The accompanying boundary conditions include the Navier slip~\citep{Navier1823,Maxwell1879} and no-penetration conditions
\begin{equation}
    \label{eq:navier_slip}
    u = \ls \pd{u}{y}, \quad v = 0,
\end{equation}
at the substrate wall $y = 0$, where the non-dimensional parameter $\ls = \lsdim/h_o$ compares the slip length, $\lsdim$, with the initial height of the film, $h_o$. The stress balance and the kinematic boundary condition,
\begin{equation}\label{eq:stressbalance_kinematic}
\tens{T} \bcdot \bm{n} + \Ca^{-1} (\nabla \bcdot \bm{n})\bm{n} = \bm{0}, \quad \bm{n} \bcdot \left( \partial_t \bm{x}_s - \bm{u} \right) = 0,
\end{equation} 
are imposed at the interface $y = h(x,t)$, where $\Ca = A/(6 \pi \sigma h_o^2) = (a/h_o)^2$ is the Capillary number, $a=[A/(6 \pi \sigma)]^{1/2}$ is the molecular length scale~\cite{de1985wetting}, $\sigma$ is the liquid-air surface tension coefficient, $\bm{x}_s = (x,h(x,t))$ is the parameterization of the free surface, and $\nabla \bcdot \bm{n} = \mathcal{C} = -\partial_x^2 h [1+(\partial_x h)^2]^{-3/2}$ is the mean curvature of the interface, with associated unit normal vector $\bm{n}$. The flow is thus governed by two dimensionless parameters, namely $\Ca$ (or equivalently $h_o/a$) and $\ls$.


\section{Leading-order lubrication models}\label{sec:LO_lubrication}

Let us now present the three leading-order lubrication models developed in previous studies. These simplified models took advantage of the existence of three possible dominant balances in the slender flow of liquid films, depending on the slip length. Specifically, these are the weak, intermediate, and strong slip limits, derived in~\cite{Sharma1996},~\cite{Munch2005}, and~\cite{kargupta2004instability}, respectively.

The weak-slip model holds when $p = \phi = O(\epsilon^{-1})$, and $\ls = O(1)$, where $\epsilon \ll 1$ measures the slenderness of the film, thus being the effect of slip a small correction to the flow driven by vdW forces,. This scaling yields the classical lubrication balance leading to a semi-parabolic profile of axial velocity, accommodating the slip condition at the substrate and the stress-free condition at the interface. In this case, the evolution of the thin film is described by the following \emph{weak-slip equation} for $h(x,t)$~\cite{Sharma1996},
\begin{equation}\label{eq:lub_weakslip}
\partial_t h  -\partial_x \left[ \frac{h^2 (h+3\ls)}{3} \partial_x \left(Ca^ {-1} \mathcal{C} + h^{-3} \right)\right] = 0,
\end{equation}
where the standard no-slip lubrication model is recovered in the regular limit $\ls \to 0$~\citep{Williams1982,craster2009dynamics}, which has been used extensively in thin-film problems~\citep{ZhangLister1999,Oron1997,Oron2000,craster2009dynamics}.

The assumption of a parabolic axial velocity profile fails when $\ls \gtrsim 1$ due to a change in the dominant balance. In particular, the strong-slip limit, $\ls\gg 1$ has associated characteristic scales $p = \phi = O(\epsilon)$ and $\ls = O(\epsilon^{-2})$, thus the flow being driven by the slip velocity at the substrate, which, at leading order, provide the \emph{strong-slip equations} describing the coupled evolution of $h(x,t)$ together with the plug-flow velocity $u(x,t)$,
\begin{subequations}
\label{eq:lub_strong}
\begin{gather}
 \partial_t h + \partial_x (h u) = 0, \\
 \frac{4}{h} \partial_x(h \partial_x u) - \partial_x \left(Ca^ {-1} \mathcal{C} +h^{-3} \right) =  \frac{u}{\ls h},
\end{gather}
\end{subequations}
which, in the limit $\ls \to \infty$, reduce to the \emph{free-film equations} derived in~\citep{Erneux1993}, and which have been used in a myriad of relevant configurations~\citep{Prevost1986,DeWit1994,Champougny2015,Thete2016,Champougny2017,Choudhury2020}.

Finally, the intermediate-slip limit can be deduced from either~\eqref{eq:lub_weakslip} or~\eqref{eq:lub_strong}. Indeed, it was shown in~\citep{Munch2005,Fetzer2005} that it can be obtained from~\eqref{eq:lub_weakslip} upon letting $t \to \ls^{-1} t$ and $\ls \to \infty$, or from~\eqref{eq:lub_strong} taking $u \to \ls u$ and $\ls \to 0$, yielding the \emph{intermediate-slip equation},
\begin{equation}\label{eq:lub_interslip}
\partial_t h  - \ls \partial_x \left[ h^2 \partial_x \left(Ca^ {-1} \mathcal{C} + h^{-3} \right)\right] = 0,
\end{equation}
with a certain range of validity around $\ls \sim 1$ to be compared against the other models below.

\section{Second-order lubrication theory}
\label{sec:parabolic_main}

A large number of higher-order lubrication models have been derived in the past to describe a wide variety of free-surface flows. In particular, regular expansions in powers of the slenderness parameter and weighted-residual approximations have been used to derive second-order models for cylindrical liquid threads~\cite{GyC,MartinezSevilla2018} and falling liquid films~\cite{Ruyer1998,Ruyer2000,Ruyer2002,Kalliadasis2011}, respectively. However, to the authors' knowledge, there is no second-order lubrication model available in the literature to describe the dynamics of ultra-thin liquid films on horizontal substrates, neither in the standard no-slip case, nor to account for wall slip. To fill this gap, here we present a second-order \textit{parabolic model}, which has $O(\epsilon^2)$ accuracy. 

The model, derived in detail in appendix~\ref{sec:parabolic}, consists of three coupled equations for the evolution of the film height, $h(x,t)$, as well as the leading- and second-order longitudinal velocities, $u_0(x,t)$ and $u_2(x,t)$, respectively, and reads
\begin{subequations}
\label{eq:lub_parabolic}
\begin{gather}
\partial_t h + \left\{h \left[u_0 \left(1 + \frac{h}{2 \ls} \right) + \frac{h^2 u_2}{3} \right] \right\}' = 0, \label{eq:kinematic_final} \\
-\left(Ca^{-1} \mathcal{C}+ h^{-3} \right)' + 3 u_0'' + 2 u_2 +  \left( \frac{(h^2 u_0)'}{h \ls} \right)' - \nonumber \\ 
\left(\frac{2(h^2 u_0')'h'}{h} + \frac{h^2}{2}(u_0''+2u_2)' - 2 (h^2 u_2)' \right)' = 0, \label{eq:xmom_final} \\
\left (\frac{u_0''}{2} + 3 u_2 \right)''- \frac{3 u_0}{\ls h^3} + \frac{3 u_0''}{h^2} + \frac{12 h' u_0'}{h^3} - \frac{6 u_2}{h^2} + \frac{3 h'^2 u_0}{\ls h^3} +\nonumber  \\
 \frac{9 u_0''}{2 \ls h} + \frac{12 h' u_0'}{\ls h^2} - \frac{3 h'^2 u_0''}{h^2} + \frac{6 h'^2 u_2}{h^2} + \frac{12 h' u_2'}{h} = 0, \label{eq:xmom_final2}
\end{gather} 
where primes indicate derivatives with respect to $x$.
\end{subequations}
The accuracy of the parabolic model~\eqref{eq:lub_parabolic} to account for the linear dynamics of the film will be assessed below.

\section{Linear stability analysis}\label{sec:LSA}

In this section we revisit the stability of slipping ultra-thin liquid films destabilized by the long-range vdW forces. Our aim is to perform a systematic comparison of the dispersion relations predicted by the different lubrication models presented in sections~\ref{sec:LO_lubrication} and~\ref{sec:parabolic_main}, with those obtained from the Stokes equations~\eqref{eq:continuity_momentum}--\eqref{eq:stressbalance_kinematic}.

\subsection{Stokes equations}\label{subsec:LSA_Stokes}

To obtain the dispersion relation $\omega = \omega(k)$ relating the longitudinal wavenumber $k$ with the temporal growth rate $\omega$, the flow is decomposed into normal modes of the form $ (u,v,p,h) = (0,0,1,1) + \delta  (\hat{u},\hat{v},\hat{p},\hat{h}) \, \exp( ikx + \omega t) $, where hats denote the eigenfunctions and $\delta\ll 1$ is the disturbance amplitude. Introducing the normal-mode decomposition into~\eqref{eq:continuity_momentum}--\eqref{eq:stressbalance_kinematic} yields the following closed-form solution,
\begin{equation}
    \label{eq:omega_stokes}
    \omega = \frac{3-\Ca^{-1} k^2}{ 2 k } 
    \frac{ 2 k \ls  \cosh (2 k)+ \sinh (2 k)  - 2 k (\ls +1) }{ 1 + (4 \ls +2) k^2+2 \ls  k \sinh (2 k)+\cosh (2 k)},
\end{equation} 
describing the temporal instability modes of the film for $0 < k < k_c = \sqrt{3 \Ca}$ and any value of $\Ca > 0$. Equation~\eqref{eq:omega_stokes} is a particular case of the dispersion relation derived in~\cite{kargupta2004instability} in the limit of negligible liquid inertia. The dispersion relation for the case with no-slip condition is recovered from~\eqref{eq:omega_stokes} by taking the limit $\ls \to 0$, yielding
\begin{equation}
    \label{eq:omega_stokes_substrate} 
    \omega_{\text{no-slip}} = \frac{3- \Ca^{-1}  k^2}{ 2 k } 
    \frac{ \sinh (2 k)  - 2 k }{ 1 + 2 k^2 +\cosh (2 k)},
\end{equation}
firstly obtained in~\cite{Jain1976}. In addition, the squeezing mode of a suspended free film is recovered in the opposite limit of perfect slip, $\ls \to \infty$, in which
\begin{equation}
    \label{eq:omega_stokes_free}
    \omega_{\text{free}} = \frac{3-\Ca^{-1}  k^2}{2 k } 
    \frac{  \cosh (2 k)-1 }{ 2k+ \sinh (2 k)},
\end{equation}
first obtained in~\cite{Maldarelli1980}. Note that, to obtain the same result as in~\cite{Maldarelli1980}, it is necessary to perform the substitution $k \to k/2$, since here we have taken the height of the film as the characteristic length scale, which is equivalent to \emph{half} the height of the equivalent free film. It is also worth pointing out that the existence of a finite slip regularizes the temporal amplification curve of the free film in the Stokes limit, in which the maximum growth rate and the associated wavenumber are $\omega^{m}_{\text{free}} \to 3/4$ and $k^m \to 0$. Indeed, the non-zero wall shear stress induced by a finite slip provides a low-wavenumber regularization mechanism alternative to liquid inertia. The temporal growth rate predicted by equation~\eqref{eq:omega_stokes} increases monotonically with $\ls$, since the fluid is able to drain faster due to the wall slippage.

\begin{figure*}
    \centering
    \includegraphics[width = 0.95\textwidth]{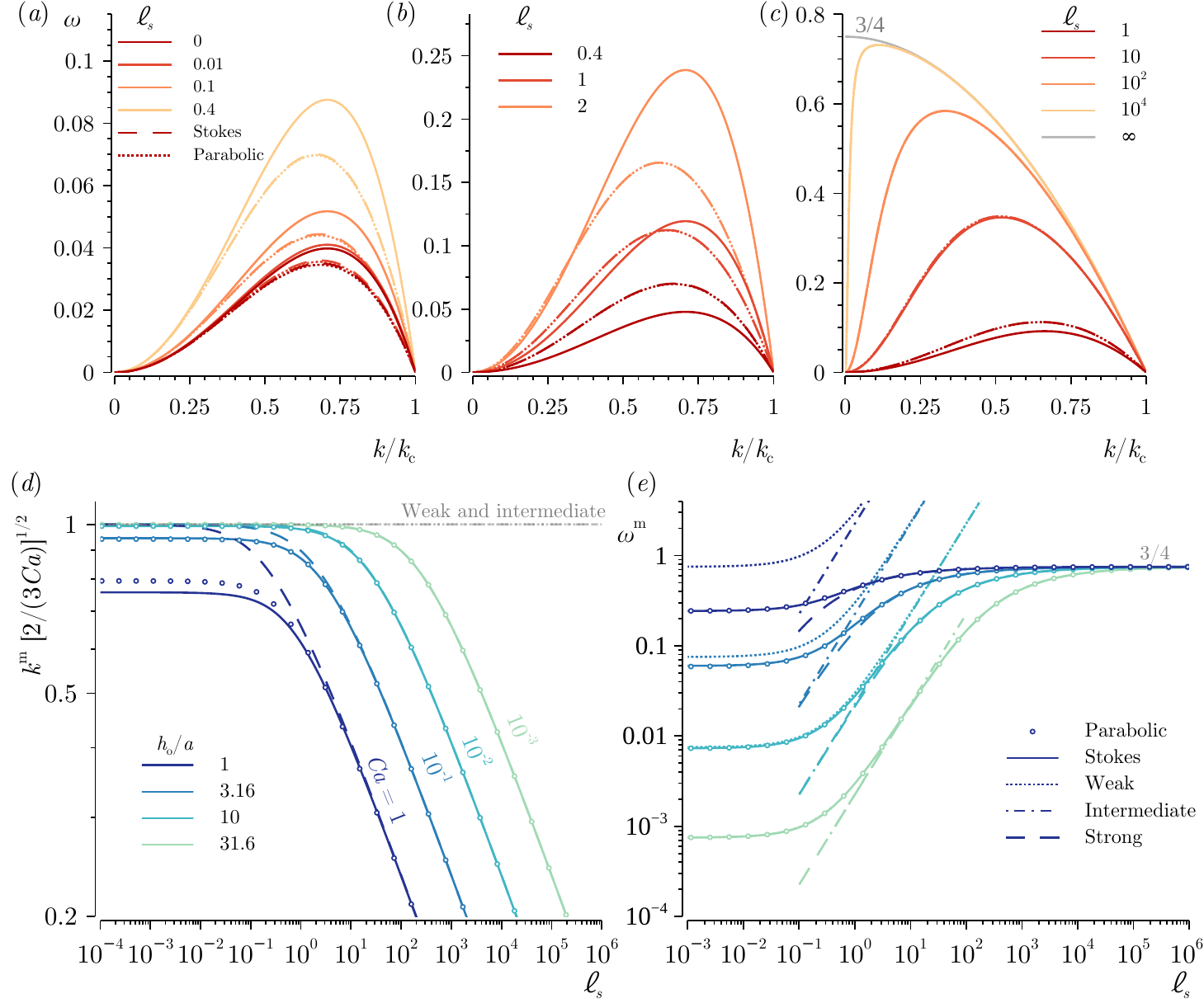}
    \caption{($a$--$c$) Amplification curves $\omega(k)$ for $\Ca = 5.31 \times 10^{-2}$ ($h_o/a = 4.34$) and several values of $\ls$ indicated in the legends for ($a$) the weak-slip, ($b$) the intermediate-slip, and ($c$) the strong-slip lubrication models~\eqref{eq:dr_lub_weak}--\eqref{eq:dr_lub_strong}, respectively (solid lines). Also shown in ($a$--$c$) are the results of the second-order parabolic model~\eqref{eq:dr_lub_parabolic} and the Stokes dispersion relation~\eqref{eq:omega_stokes} (dashed and dotted lines, respectively). Note that $k$ is scaled with the cut-off wavenumber $k_c = \sqrt{3 \Ca}$. ($d$) Wavenumber of maximum amplification, $k^m$, scaled with $(3\Ca/2)^{1/2}$, and ($e$) corresponding maximum growth rate, $\omega^m$, as functions of the dimensionless slip length $\ls$, obtained from the five frameworks as indicated in the legend. \label{fig:fig2}}
\end{figure*}

\begin{figure*}
    \centering
    \includegraphics[width = 0.95\textwidth]{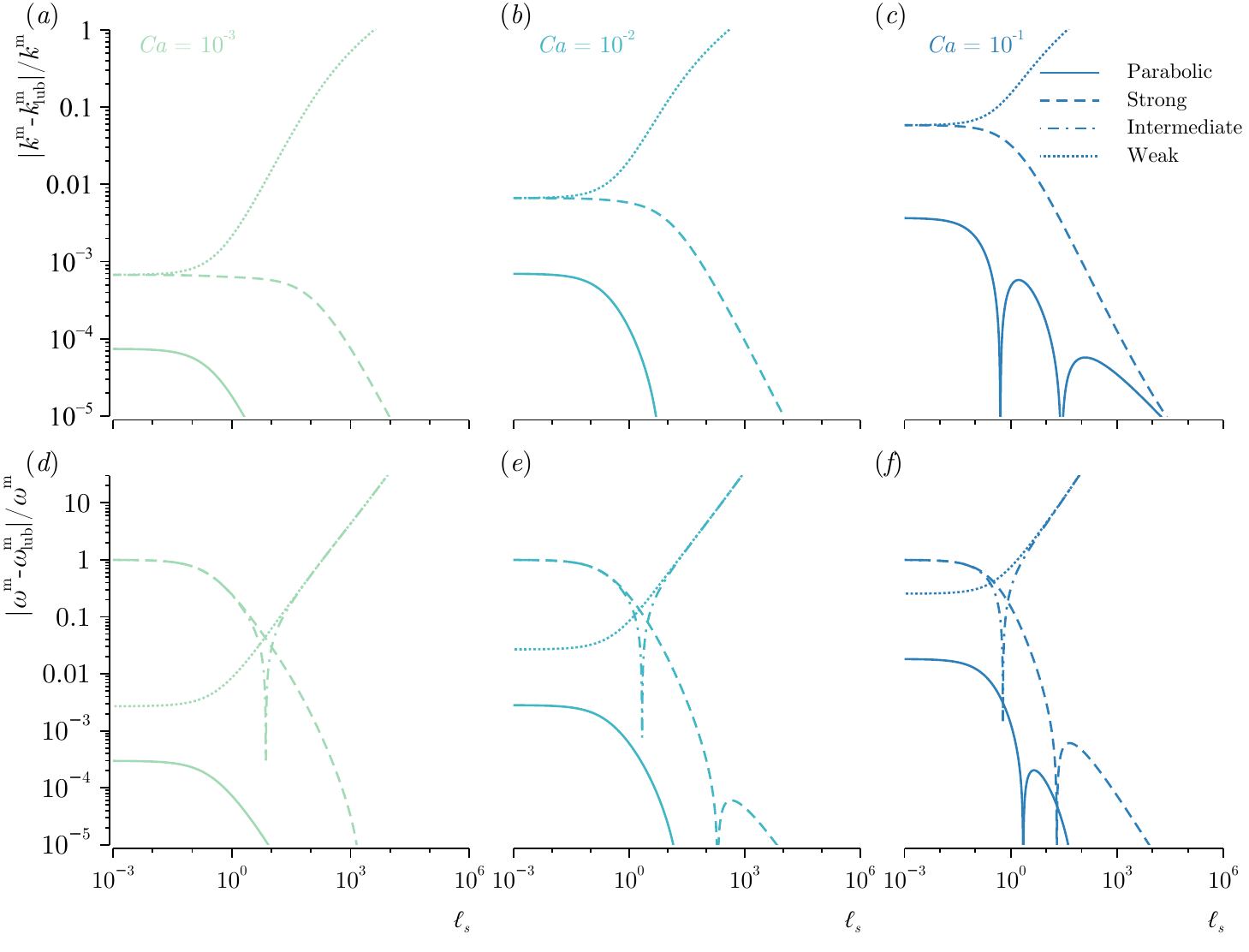}
    \caption{The relative errors in ($a$--$c$) wavenumber and ($b$--$f$) growth rate of most unstable perturbations defined for the four lubrication models, indicated in the legend, with respect to the Stokes equations for $\Ca = 10^{-3}$ in ($a$,$d$), $10^{-2}$ in ($b$,$e$), and $10^{-1}$ in ($c$,$f$).}
    \label{fig:fig3}
\end{figure*}

\subsection{Lubrication theory}\label{subsec:LSA_Lub}

Following the same procedure explained in~\S\ref{subsec:LSA_Stokes}, we obtain the corresponding dispersion relations of the three leading-order lubrication approximations~\eqref{eq:lub_weakslip}--\eqref{eq:lub_interslip}, and the one associated with the parabolic model~\eqref{eq:lub_parabolic}, namely
\begin{subequations}
\begin{gather}
\omega_{\text{w}} = \frac{k^2(3-\Ca^{-1} k^2)(1+3 \ls)}{3}, \label{eq:dr_lub_weak} \\
\omega_{\text{i}} = \ls k^2(3- \Ca^{-1} k^2), \label{eq:dr_lub_intermediate} \\
\omega_{\text{s}} = \frac{k^2(3-\Ca^{-1} k^2)}{4 k^2+\ls^{-1}}, \label{eq:dr_lub_strong}\\
\omega_{\text{p}} = \frac{k^2(3-\Ca^{-1} k^2)[4 +(6+k^2)^2\ls/3]}{12 + k^2 \, \left\{ 24 + 48 \ls  + k^2[ 4(4+k^2)\ls - 3]\right\}}, \label{eq:dr_lub_parabolic}
\end{gather}
\end{subequations}
where the subscripts denote the \emph{w}eak-slip (hereafter WSM), \emph{i}ntermediate-slip (hereafter ISM), \emph{s}trong-slip (hereafter SSM) and parabolic models, respectively. Note that the growth rate predicted by the WSM recovers the particular case of a perfect no-slip condition at the substrate, in which $\omega_{\text{w}} \to k^2(3-\Ca^{-1} k^2)/3$~\citep{Vrij1966,Williams1982} as $\ls \to 0$, but fails catastrophically for $\ls \gtrsim 1$. In the opposite limit, the growth rate obtained from the SSM recovers the free-film lubrication dispersion relation, $\omega_{\text{s}} \to (3-\Ca^{-1} k^2)/4$ as $\ls \to \infty$~\citep{Erneux1993}, but vanishes for $\ls \to 0$. The ISM is seen to fail in both limits. Most importantly, the second-order parabolic model developed herein is regular for all the values of $\ls$, and perfectly captures both the no-slip and the free-film limits.

Figures~\ref{fig:fig2}($a$-$c$) show the amplification curves obtained from the different models for $\Ca = 5.31 \times 10^{-2}$ ($h_o/a = 4.34$, a similar value to the one reported by~\citep{becker2003complex,Reiter2001}), and different values of the dimensionless slip length $\ls$ indicated in the legends. Note that, for convenience, $k$ has been rescaled with the cut-off wavenumber $k_c$ in figures~\ref{fig:fig2}($a$--$c$), so that the effective length scale is $h_o^2/(\sqrt{3} a)$ instead of $h_o$ in these plots. The amplification curves obtained from the WSM, ISM and SSM are shown as solid lines in figures~\ref{fig:fig2}($a$--$c$), respectively, while the results of the parabolic model~\eqref{eq:dr_lub_parabolic}, and the Stokes equations~\eqref{eq:omega_stokes}, are also shown for several cases using dashed and dotted lines, respectively. Figures~\ref{fig:fig2}($a$--$c$) reveal that, for $\Ca = 5.31 \times 10^{-2}$, the results of the parabolic model are almost identical as those obtained from the Stokes equations for any value of $\ls$. As revealed by figure~\ref{fig:fig2}($a$), the WSM exhibits relative differences of about 10\%  with respect to the Stokes results, even for $\ls = 0$, and introduces 100\% errors when $\ls \sim 5$. The ISM is seen to have large associated relative errors even for values of $\ls \sim 1$, as deduced from figure~\ref{fig:fig2}($b$). Finally, figure~\ref{fig:fig2}($c$) shows that the SSM performs remarkably well, even for values of $\ls \sim 1$.

To quantify the performance of the lubrication models, figures~\ref{fig:fig2}($d$,$e$) display the wavenumber for maximum amplification, $k^m$, and the corresponding growth rate, $\omega^m$, as functions of $\ls$, for several values of $\Ca$ indicated near each curve. In addition, the relative errors in $k^m$ and $\omega^m$, measured with respect to the Stokes solution, are shown in figures~\ref{fig:fig3}($a$-$f$). Notice that $k^m$ and $\omega^m$ can be obtained analytically from the lubrication models, yielding
\begin{subequations}
\begin{gather}
k_{\text{w}}^m = \sqrt{\frac{3 \Ca}{2}} = \frac{k_c}{\sqrt{2}}, \quad \omega_{\text{w}}^m =\frac{3 \Ca(1+3 \ls)}{4}, \label{eq:max_lub_weak} \\
k_{\text{i}}^m = \sqrt{\frac{3 \Ca}{2}} = \frac{k_c}{\sqrt{2}}, \quad \omega_{\text{i}}^m =\frac{9 \Ca \ls}{4}, \label{eq:max_lub_intermediate} \\
k_{\text{s}}^m = \frac{1}{2}\sqrt{\frac{ (1 + 12 \ls \Ca)^{1/2} - 1}{\ls}}, \quad \omega_{\text{s}}^m =\frac{1+6 \ls \Ca - \sqrt{1+ 12 \Ca \ls}}{8 \Ca \ls}, \label{eq:max_lub_strong}
\end{gather}
\end{subequations}
whereas the exact expressions for $k^m$ and $\omega^m$ deduced from the parabolic model are implicit, and are not shown here for conciseness. As shown in figures~\ref{fig:fig2}($d$,$e$) and~\ref{fig:fig3}($a$--$f$), the SSM provides a good approximation to the Stokes equations for $\ls \gtrsim 1$ and $\Ca \lesssim 1$. Note that the maximum admissible value of $\Ca$ under realistic conditions is $\Ca \approx 1$, since it corresponds to $h_o\approx a$, i.e. a {\em molecularly thin} film. The WSM is seen to perform reasonably well for $\ls \lesssim 10$ and $\Ca \lesssim 10^{-2}$, where the error in $\omega^m$, which in this case is the most restrictive one, is always below about 10$\%$. The ISM has small relative errors within an extremely narrow window of slip lengths for any value of $\Ca$, as shown in figures~\ref{fig:fig2}($e$) and~\ref{fig:fig3}($d$--$f$), what strongly compromises its practical use. Finally, the relative errors in $\omega^m$ and $k^m$ associated to the parabolic model are always below 1$\%$ and 4$\%$, respectively, for arbitrary values of $\ls$ and $\Ca$.

We will now take into account that the unstable range of wavenumbers is $0\leq k\leq k_c$, with $k_c=(3\Ca)^{1/2}$. Since $k_c\leq 1$ for $\Ca<1/3$, further analytical insight into the different lubrication models can be gained by performing small-$k$ expansions of the difference between their respective dispersion relations~\eqref{eq:dr_lub_weak}--\eqref{eq:dr_lub_parabolic}, and the Stokes result~\eqref{eq:omega_stokes}, to yield
\begin{subequations}
\begin{gather}
\omega-\omega_{\text{w}} =  -\frac{9+15\ls(3+4\ls)}{5}k^4 + O(k^6), \label{eq:weak_smallk} \\
\omega-\omega_{\text{i}} =k^2 - \left(\frac{1}{3 Ca}+ \frac{9+15\ls(3+4\ls)}{5} \right)k^4 + O(k^6), \label{eq:intermediate_smallk} \\
\omega- \omega_{\text{s}} = k^2 - \left(\frac{1}{3 Ca}+ \frac{9\ls(1+5\ls)}{5} \right)k^4 + O(k^6), \label{eq:strong_smallk}\\
\omega-\omega_{\text{p}} = \frac{k^4}{5}- \left(\frac{1}{15 Ca} + \frac{27 + 70 \ls}{28} \right)k^6 + O(k^8), \label{eq:parabolic_smallk}
\end{gather}
\end{subequations}
which are formally valid for $\Ca<1/3$. The errors associated with the weak and parabolic lubrication models are $O(k^4)$, although in the latter case the slip length appears only at $O(k^6)$ and, in addition its $k^6$ coefficient is much smaller than that of the weak model (not shown). Moreover, in the no-slip limit, $\ls \to 0$, the $k^4$ coefficient is nine times smaller in the case of the parabolic model, what explains its much better performance when $\Ca \to 1$.

The two key variables that can be extracted from the linear stability analysis to compare with the experiments are the film rupture time, $t_R$, and the wavenumber of maximum temporal amplification, $k^m$, which allows to estimate the characteristic length scale of the dewetting pattern as $\lambda \sim 2\pi/k^m$. Thus, we will finally assess the predictions of the film rupture time, $t_R$, provided by the different models. To that end, we take into account that when the initial disturbance amplitude $\delta\ll 1$, the thinning of the film follows the linearized dynamics during most of its time evolution~\citep{morenobozaetal2020}, $\hmin(t) =1 + \delta \exp(\omega^m t)$, where $\hmin(t)$ is the minimum film thickness, leading to the estimation $t_R = \ln(\epsilon^{-1})/\omega^m$. The WSM, ISM and SSM models yield closed analytical expressions,
\begin{subequations}
\begin{gather}
\frac{t_R^{\text{w}}}{\ln(\epsilon^{-1})} = \frac{4}{3 \Ca (1+ 3 \ls)},\label{eq:tr_weak}\\
\frac{t_R^{\text{i}}}{\ln(\epsilon^{-1})} = \frac{4}{9 \Ca \ls}, \label{eq:tr_inter}\\
\frac{t_R^{\text{s}}}{\ln(\epsilon^{-1})} = \frac{8 \Ca \ls}{1+6 \Ca \ls - \sqrt{1+12 \Ca \ls}} = \nonumber\\
\frac{4}{3} +\frac{4}{3 \sqrt{3 \Ca \ls}} - \frac{2}{9 \Ca  \ls} + O(\Ca^{-3/2} \ls^{-3/2}). \label{eq:tr_strong}
\end{gather}
\end{subequations}
To obtain explicit equations for $t_R$ in the case of the parabolic model and the Stokes equations, it is first necessary to perform a small-$k$ expansion, and then expand in powers of $\Ca$ (i.e. inverse powers of $h_o/a$). In the weak-slip limit $\ls \ll 1$, the resulting predictions for the rupture time are
\begin{subequations}
\begin{gather}
\frac{t_R^{\text{p}}}{\ln(\epsilon^{-1})} = \frac{4}{3 \Ca (1+ 3 \ls)} + \frac{4 + 6 \ls (3 + 4 \ls)}{(1 + 3 \ls)^2} + O(\Ca), \label{eq:tr_parabolic}\\
\frac{t_R^{\text{Stokes}}}{\ln(\epsilon^{-1})} = \frac{4}{3 \Ca (1+ 3 \ls)} + \frac{6 [3 + 5 \ls (3 + 4 \ls)]}{5 (1 + 3 \ls)^2}+ O(Ca), \label{eq:tr_stokes}
\end{gather}
\end{subequations}
while, in the strong-slip limit $\ls \gg 1$,
\begin{equation}
\frac{t_R^{\text{Stokes}}}{\ln(\epsilon^{-1})} = \frac{t_R^{\text{p}}}{\ln(\epsilon^{-1})} = \frac{4}{3} + \frac{4}{3 \sqrt{3 Ca \ls}} + \frac{3+Ca}{9 \Ca \ls} + O(\ls^{-3/2}), \label{eq:tr_bigls}
\end{equation}
which yields exactly the same result in both frameworks, and slightly different from the SSM~\eqref{eq:tr_strong}. In particular, both equations,~\eqref{eq:tr_strong} and~\eqref{eq:tr_bigls}, are independent of $\Ca$ at leading order, which can be obtained in the free-film limit, $\ell_s \gg 1$. Moreover, note that~\eqref{eq:tr_weak},~\eqref{eq:tr_parabolic} and~\eqref{eq:tr_stokes} are identical at leading order, and recover the result obtained in~\cite{morenobozaetal2020} for a non-slipping ultra-thin liquid film in the Stokes regime.

\section{Conclusions}\label{sec:conclusions}

A comprehensive analysis of the linear stability properties of slipping ultra-thin films has been carried out in the limit of negligible inertia. The three lubrication models available in the literature to account for wall slip during the thinning of the film, derived for weak~\cite{Sharma1996}, intermediate~\cite{Munch2005}, and strong~\cite{kargupta2004instability} slip regimes, have been critically assessed by a systematic comparisons of their predictions with those stemming from a linear stability analysis of the fully two-dimensional Stokes equations, with emphasis on the two key magnitudes predicted by the stability analysis, namely the maximum growth rate and the associated optimal wavenumber. 

The weak and strong lubrication models show a good quantitative performance for small and large slip lengths, respectively. In the case of the weak-slip model, the agreement requires the film to be initially slender, what is guaranteed for small enough values of the Capillary number, $\Ca \ll 1$ or, equivalently, for large enough values of the initial film thickness compared with the molecular length scale, $h_o \gg a$. In the case of the strong-slip model, since the optimal wavenumber $k^m \to 0$ as $\ls \gg 1$, the agreement is independent of $\Ca$, and corresponds to the free-film limit. More importantly, the intermediate-slip model, which is routinely used~\citep{Peschka2019}, has an extremely narrow window of validity of $O(1)$ slip lengths, a conclusion that should be taken into account in future studies of slipping films.

Due to the limitations of the leading-order lubrication models, we have developed a second-order parabolic lubrication model, whose performance in the linear regime is on par with the full Stokes equations for arbitrary slip lengths. The latter result suggests that the new second-order lubrication model could be helpful in future numerical analyses of slipping ultrathin films. In particular, the parabolic model requires integrating three coupled nonlinear partial differential equations for the film height and the leading- and second-order velocities, arising in a regular asymptotic expansion in which the small parameter is the transverse coordinate~\cite{GyC,MartinezSevilla2018}. 


\section*{Acknowledgments}

This research was funded by the Spanish MINECO, Subdirecci\'on General de Gesti\'on de Ayudas a la Investigaci\'on, through project DPI2015-71901-REDT, and by the Spanish MCIU-Agencia Estatal de Investigaci\'on through project DPI2017-88201-C3-3-R, partly financed through FEDER European funds. A.M.-C. also acknowledges support from the Spanish MECD through the grant FPU16/02562.

\section*{Data availability}

The data that support the findings of this study are available from the corresponding author upon reasonable request.

\begin{appendix}
\section{Detailed derivation of the parabolic model}\label{sec:parabolic}

Following previous studies~\cite{GyC,MartinezSevilla2018}, the second-order lubrication model is derived by introducing the following rescaled characteristic variables
\begin{subequations}
\begin{gather}
x_c = y_c = \frac{h_o}{\epsilon}, \quad t_c = \frac{6 \pi \mu h_o^3}{\epsilon A},\label{eq:scales_parabolic1}\\
u_c = v_c = \frac{A}{6 \pi \mu h_o^2}, \quad h_c = h_o, \quad p_c = \frac{A}{6 \pi h_o^3},\label{eq:scales_parabolic2}
\end{gather}
\end{subequations}
where $\epsilon = h_o/L \ll 1$, with $L$ being a characteristic longitudinal length. We expand all the flow variables in $y \sim \epsilon$ as follows
\begin{subequations}
\begin{gather}
u(x,y,t) = u_0(x,t) + \frac{u_0(x,t)}{\ls}y + u_2(x,t)y^2 + u_3(x,t)y^3 +..., \label{eq:exp_u} \\
v(x,y,t) = -u_0'(x,t)y - \frac{u_0'(x,t)}{2\ls}y^2 - \frac{u_2'(x,t)}{3}y^3-..., \label{eq:exp_v} \\
p(x,y,t) = p_0(x,t) + p_1(x,t)y + p_2(x,t)y^2 + ..., \label{eq:exp_p}
\end{gather}
\end{subequations}
where the expansions of $u$ and $v$ satisfy the continuity equation~\eqref{eq:continuity_momentum}, as well as the slip and no penetration boundary conditions at the wall given by equation~\eqref{eq:navier_slip}.

At $O(\epsilon^2)$ the kinematic condition at $y = \epsilon h$ and the hierarchy of equations obtained from the axial momentum equation read
\begin{subequations}
\begin{gather}
\partial_t h + \left[h \left(u_0 \left( 1 + \frac{h }{2 \ls} \right) + \frac{h^2 u_2}{3} \right) \right]' + O(\epsilon^4) = 0, \label{eq:kinematic_p} \\
-\left(p_0+ h^{-3} \right)' + \epsilon (u_0'' + 2 u_2) + O(\epsilon^2) = 0, \label{eq:xmom_p0} \\
-p_1' + \epsilon \left(\frac{u_0''}{\ls} + 6 u_3\right)+ O(\epsilon^2) = 0, \label{eq:xmom_p1} \\
-p_2' + \epsilon (u_2'' + 12 u_4)+ O(\epsilon^2) = 0, \label{eq:xmom_p2}
\end{gather}
\end{subequations}
where the functions $p_1(z,t)$ and $p_2(z,t)$ can be determined from the corresponding orders of the $y$-momentum equation, which are then substituted back into~\eqref{eq:xmom_p1} to determine $u_3(x,t)$, yielding
\begin{subequations}
\begin{gather}
p_1(x,t) = -\frac{\epsilon u_0'}{\ls} + O(\epsilon^2), \label{eq:p1} \\
p_2(x,t) = -\frac{\epsilon}{2}(u_0''+2 u_2)' + O(\epsilon^2), \label{eq:p2} \\
u_3(x,t) = - \frac{u_0''}{3 \ls}, \label{eq:u3}
\end{gather}
\end{subequations}
as well as the following equation for~\eqref{eq:xmom_p2},
\begin{equation}
\left( u_0''/2 + 2 u_2 \right)'' + 12 u_4 + O(\epsilon) = 0, \label{eq:xmom_p2_new}
\end{equation}
To obtain a closed system for $h(x,t)$, $u_0(x,t)$ and $u_2(x,t)$, we need to obtain $p_0(x,t)$ in~\eqref{eq:xmom_p0}, and $u_4(x,t)$ in~\eqref{eq:xmom_p2_new}, from the normal and tangential stress balances at $O(\epsilon^2)$,
\begin{align}\label{eq:p_0}
& p_0(x,t) =  Ca^{-1} \mathcal{C}-2 \epsilon u_0' - \epsilon^2  \frac{(h^2 u_0)'}{h \ls} + \nonumber & \\
& \epsilon^3 \left(\frac{2(h^2 u_0')'h'}{h} + \frac{h^2}{2}(u_0''+2u_2)' - 2 (h^2 u_2)' \right) + O(\epsilon^4)
\end{align}
\begin{align}\label{eq:u_4}
& \epsilon^3 u_4(x,t) =  - \frac{u_0}{4 h^3 \ls} + \epsilon \left(\frac{u_0''}{4 h^2} + \frac{h' u_0'}{h^3} - \frac{u_2}{2 h^2}  \right) + \nonumber & \\
& \epsilon^2 \left( \frac{h'^2 u_0}{4 \ls h^3} + \frac{3 u_0''}{8 \ls h} + \frac{h' u_0'}{\ls h^2} \right) + \epsilon^3 \left( \frac{h'^2 u_2}{2 h^2} - \frac{h'^2 u_0''}{4 h^2} + \frac{u_2''}{12} + \frac{h' u_2'}{h} \right),
\end{align}
respectively. Introducing~\eqref{eq:p_0} into~\eqref{eq:xmom_p0} and~\eqref{eq:u_4} into~\eqref{eq:xmom_p2_new}, and eliminating $\epsilon$ upon the substitution $x \to \epsilon x$, $t \to \epsilon t$, $u_{2j} \to u_{2j}/\epsilon^{2j}$ ($j = 0,1$) and $\ls \to \epsilon \ls$, yields the parabolic model~\eqref{eq:lub_parabolic} as a closed system to determine $h(x,t)$, $u_0(x,t)$ and $u_2(x,t)$.



\end{appendix}


%

\end{document}